\pdfoutput=1

\documentclass[11pt]{article}

\usepackage[]{acl}

\usepackage{times}
\usepackage{latexsym}

\usepackage[T1]{fontenc}

\usepackage[utf8]{inputenc}

\usepackage{microtype}
\usepackage[skins]{tcolorbox}

\usepackage{amsfonts}
\usepackage{amsmath}
\usepackage{multirow,makecell}
\usepackage{booktabs}
\usepackage{graphicx,subfigure}
\newcommand{\hide}[1]{} 
\def\code#1{\texttt{#1}}

\usepackage{adjustbox}
\usepackage{balance}
\usepackage{array}
\newcolumntype{H}{>{\setbox0=\hbox\bgroup}c<{\egroup}@{}}

\usepackage{acronym}
\usepackage{enumitem}

\newcommand{\commentout}[1]{}
\newcommand{\eqnref}[1]{Eq.~\ref{#1}}

\newcommand{\secref}[1]{Section~\ref{#1}}
\newcommand{\figref}[1]{Figure~\ref{#1}}

\newcommand{\tblref}[1]{Table~\ref{#1}}

\acrodef{LLM}{large language model}
\acrodef{PRP}{pairwise-ranking prompting}

\title{Consolidating Ranking and Relevance Predictions \\ of Large Language Models through Post-Processing}
\author{
    Le Yan, Zhen Qin, Honglei Zhuang, Rolf Jagerman, \\
    \textbf{Xuanhui Wang, Michael Bendersky, Harrie Oosterhuis} \\
    Google Research\\
    \texttt{\{lyyanle,zhenqin,hlz,jagerman,xuanhui,bemike,harrie\}@google.com}
}

\begin{document}

\maketitle
\begin{abstract}
The powerful generative abilities of \acp{LLM} show potential in generating relevance labels for search applications.
Previous work has found that directly asking about relevancy, such as ``\emph{How relevant is document A to query Q?}", results in sub-optimal ranking.
Instead, the \ac{PRP} approach produces promising ranking performance through asking about pairwise comparisons, e.g., ``\emph{Is document A more relevant than document B to query Q?}".
Thus, while \acp{LLM} are effective at their ranking ability, this is not reflected in their relevance label generation.

In this work, we propose a post-processing method to consolidate the relevance labels generated by an \ac{LLM} with its powerful ranking abilities. Our method takes both \ac{LLM} generated relevance labels and pairwise preferences.
The labels are then altered to satisfy the pairwise preferences of the \ac{LLM}, while staying as close to the original values as possible.
Our experimental results  indicate that our approach effectively balances label accuracy and ranking performance.
Thereby, our work shows it is possible to combine both the ranking and labeling abilities of \acp{LLM} through post-processing.




\end{abstract}

\acresetall
\section{Introduction}

Generative large language models (LLM) have shown significant potential on question answering and other conversation-based tasks~\cite{gpt4,palm2} owing to their extraordinary generative abilities and natural language understanding capabilities.
Naturally, previous work has further investigated the application of LLMs to other areas, including search and recommendation tasks~\cite{zhu2023large,wu2023survey}.
The goal here is to rank items according to their relevance to a certain query.
Generally, existing approaches have applied LLMs to this task in two different ways:
First, as pseudo-raters, LLMs are asked to simulate human raters by generating a relevance label for each query-document pair~\cite{liang2022holistic}, for example, through prompts such as ``\emph{How relevant is document A to query Q?}"
Secondly, an LLM can also be asked directly about the ordering of documents for a query, for example, the pairwise-ranking-prompting (PRP) method~\cite{qin2023large} uses a prompt like ``\emph{Is document\ A more relevant than document\ B to query Q?}"
Alternatively, LLMs can be asked to generate the entire ranking through a prompt like ``\emph{Rank the following documents by their relevance to query Q: document A, document B, document C, etc.}''~\cite{baidugpt}
Thus, there are several distinct modes by which LLMs can be used for ranking purposes, which provide different kinds of output.

Each mode of applying LLMs to ranking tasks offers distinct advantages in terms of performance and efficiency.
The pseudo-rater mode is currently favored in LLM applications within ranking systems due to its simplicity and high efficiency
~\cite{liang2022holistic,sachan2022improving,thomas2023large}.
Given the high costs associated with deploying or training LLMs for high-throughput applications like search and recommendations, it is only efficiently feasible to use LLMs as pseudo-raters to label a fraction of raw data in zero-shot or few-shot fashion as a replacement of more expensive human raters. 
However, the general LLMs are not tuned to generate meaningful ranking scores, as a result, there is still an apparent gap between state of the art (SOTA) ranking performance and the performance reached when leveraging LLM pseudo-labels for model training~\cite{thomas2023large}. 

In parallel to exploring the costly fine-tuning of LLMs as ranking specialists~\cite{monot5,zhuang2023rankt5}, previous work has also investigated the direct ranking modes of LLMs.
Some of these direct ranking modes, such as PRP~\cite{qin2023large}, can reach SOTA ranking performance that is on-par with LLMs finetuned for ranking.
Moreover, PRP enables open-source (OSS) LLMs to outperform the largest commercial models like GPT-4~\cite{gpt4}.
However, document scoring by PRP solely considers the resulting order of the candidate list, and thus, the absolute values of scores are meaningless.
This makes PRP results unsuitable to be directly used as pseudo-labels.
For example, the PRP ranking score of a fair candidate in the list of only poor candidates would be comparable to that of a good candidate in the list of strong competing candidates, see example in \figref{fig:rankingrater}. 
How to effectively combine these two modes to consolidate ranking and relevance predictions of LLMs remains an essential challenge in applying LLMs to real world main stream applications.

In this work, we study post-processing methods to do the consolidation, especially for the case when we have no human labelled data. 
We first define the problem in LLM ranking in~\secref{sec:problem}, and propose our post-processing methods to consolidate LLM predictions for unlabelled data in~\secref{sec:methods}. 
We discuss our experiments on public ranking datasets in~\secref{sec:exp} and show our methods could approach the state of the art ranking performance with minimal tradeoff in relevance prediction performance in~\secref{sec:results}.
Our contributions include: 
\begin{itemize}[leftmargin=*]
    \item The first systematic study on the tradeoff between ranking and relevance predictions of LLMs.
    \item A ranking-aware pseudo-rater pipeline with a novel post-processing method using constrained regression to combine both PRP ranking and LLM relevance generation.
    \item Extensive experimental study on public ranking datasets that demonstrates the effectiveness of our proposed methods.
\end{itemize}
\section{Related Work} \label{sec:related}
The strong capability of LLMs in textual understanding has motivated numerous studies leveraging LLM-based approaches for textual information retrieval~\cite{bonifacio2022inpars,tay2022transformer,jagerman2023query}. 
Before the generative LLM era, the focus was more on finetuning pre-trained language models (PLMs) such as T5~\cite{monot5,zhuang2023rankt5} or BERT~\cite{nogueira2019passage} for the supervised 
learning to rank problem~\cite{8186875,dasalc}, which becomes less feasible with larger generative LLMs. 
Two popular methods—-relevance generation~\cite{liang2022holistic,zhuang2023beyond} and query generation~\cite{sachan2022improving}-—aim to generate per-document relevance scores or retrieval queries using generative LLMs. 
These methods are also termed pointwise approaches for ranking.
More recent works~\cite{baidugpt,jimmygpt, pradeep2023rankvicuna, tang2023found} explore listwise ranking generation approaches by directly inserting the query and a list of documents into a prompt. 
Pairwise order generation through pairwise prompts~\cite{qin2023large} turns out to be very effective for ranking purposes, especially for moderated-sized LLMs. 
However, none of these ranking approaches using generative LLMs attempt to consolidate the results with relevance generation.

Previous works on non-LLM neural rankers~\cite{yan2022scale,bai2023regression} focus on balancing or aligning regression with ranking objectives during the model training, which is unfortunately not feasible for LLMs using zero-shot or few-shot prompting. 
Post-processing methods that calibrate model predictions using some validation data could be potentially applicable.
Originally developed for classification model calibration~\cite{Menon:ICML12}, these methods include parametric approaches like Platt scaling~\cite{PlattScaling:1999} for binary classification; piecewise linear transformation~\cite{ravina2021distilling} for regression; and non-parametric approaches like isotonic regression~\cite{Menon:ICML12,Zadrozny:Elkan:KDD02}, histogram binning, and Bayesian binning~\cite{Zadrozny:Elkan:KDD01,naeini2015obtaining}. 
But how effectively these post-processing approaches could be extended to LLM-based ranking and relevance predictions has not been well studied in existing literature. 

\section{Problem Formulation} \label{sec:problem}
We formulate the problem of consolidating ranking and relevance predictions within this framework. 
Given a set of queries, for each query $q$, we have a  set of corresponding candidate documents $\{d\}_q$, and their ground truth labels, $\{y\}_q$, as their relevance evaluations, such as graded relevance.
Our first goal is to predict the relevance labels based on the content of each corresponding candidate.
Our second goal is to predict a ranked list of candidates, and we use $\{r\}_q$ to denote the rank of each candidate in this predicted ranking.
The predicted ranking is optimal when the ranks align with the order of the relevance labels: $r_i\leq r_j$ if $y_i\geq y_j$ for any pair of candidates $(d_i, d_j)$ belonging to the same query $q$.
Taken together, our overall task is to optimize LLM predictions for both relevance estimation and ranking performance. 

\begin{figure*}[htbp]
  \centering
  \includegraphics[width=.22\linewidth]{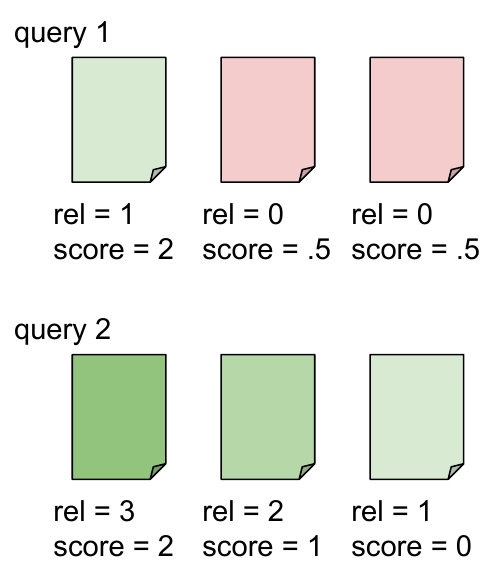}\qquad\includegraphics[width=.7\linewidth]{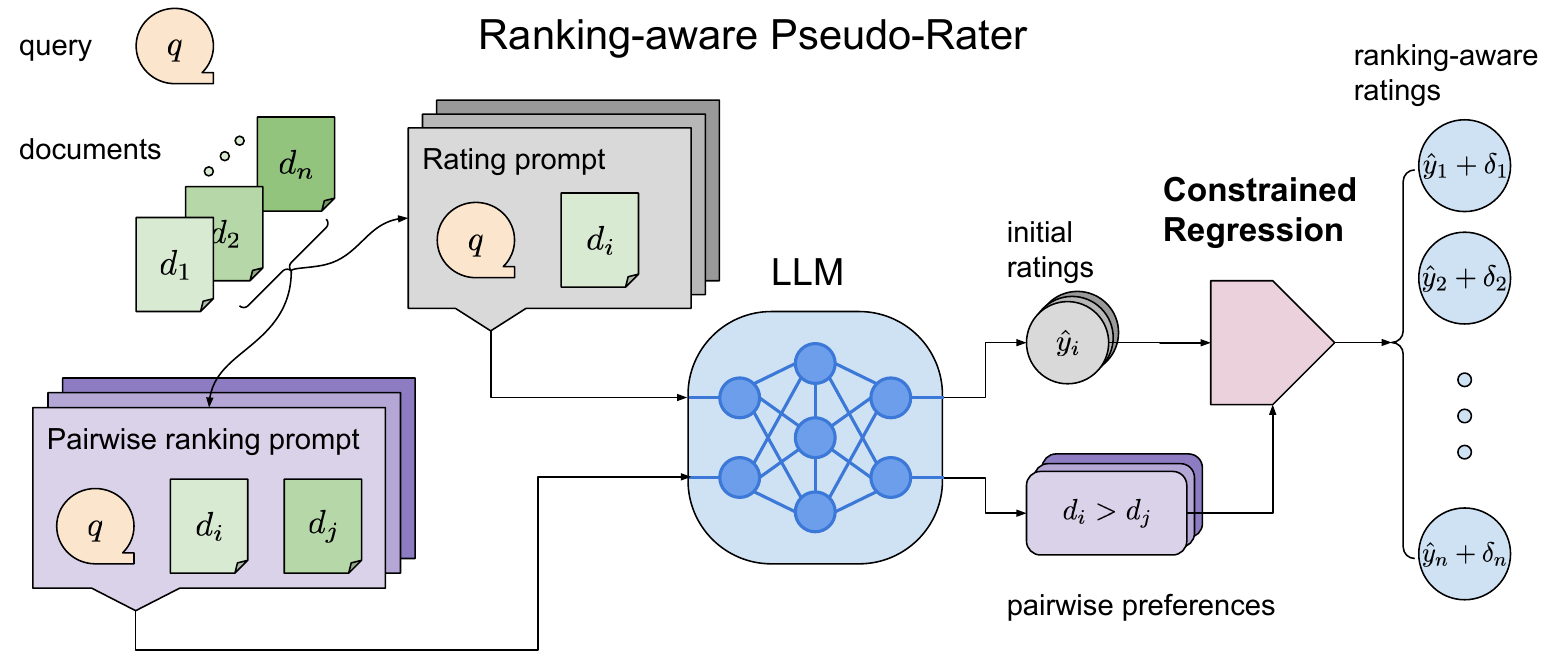}
  \caption{Left: Example of PRP scores not calibrated over different queries. Right: Illustration of the ranking-aware pseudo-rater pipeline that generates ranking-aware ratings with LLMs from the input query and list of candidate documents.}
  \label{fig:rankingrater}
\vspace{-0.5cm}
\end{figure*}

\subsection{Relevance Prediction}
For this purpose, in this work, we consider real-number predictions, i.e., $\hat{y}_i\in\mathbb{R}$, as the relevance pseudo-labels for query-document pairs. Such pointwise real-number ratings can be averages over the annotations of multiple human raters. 
For LLM-based raters, pseudo-labels can be obtained from the average rating of raters with discrete output space~\cite{thomas2023large} or from finer-grained rating generation~\cite{zhuang2023beyond}, or directly leveraging the token probabilities to formulate the relevance predictions if available in the generative LLMs~\cite{liang2022holistic}.

In specific, we use LLM as a rater to generate ``Yes'' or ``No'' to answer the question ``does the passage answer the query?'' for each query-document pair. See Appendix~\ref{sec:prompt_relevance} for the prompt. We obtain the generation probabilities $P_i(\textrm{Yes})$, $P_i(\textrm{No})$ and take
\begin{equation}
\label{eq:relevance}
    \hat{y}_i = \frac{P_i(\textrm{Yes})}{P_i(\textrm{Yes})+P_i(\textrm{No})}
\end{equation}
as the normalized relevance prediction: $\hat{y}_i=1$ for the most relevant document and $\hat{y}_i=0$ for the least.

To evaluate the relevance prediction performance of $\{\hat{y}\}_q$, we consider 
the mean squared error (MSE):
\begin{equation}
\label{eq:mse}
    {\rm MSE}(\{y\}_q, \{\hat{y}\}_q) = \frac{1}{|\{d\}_q|}\sum_{i \in \{d\}_q}(\hat{y}_i - y_i)^2,
\end{equation}
%
as well as the empirical calibration error (ECE)~\cite{naeini2015obtaining, guo2017calibration}:
\begin{equation}
\label{eq:ece}
    {\rm ECE}_q = \frac{1}{|\{d\}_q|}\sum_{m=1}^{M}\left|\sum_{i\in B_m}y_i-\sum_{i\in B_m}\hat{y}_i\right|,
\end{equation}
where we group candidates of each query into $M$ successive bins of model score-sorted results $B_m$, and ${|\{d\}_q|}$ gives the size of candidate documents to query $q$. Compared to MSE, ECE is more sensitive to the distribution divergence between predictions and ground truth labels due to binning.


\subsection{Ranking Prediction}
In the pairwise ranking prompting (PRP) mode, LLMs generate pairwise preferences: for any two documents $d_1$ and $d_2$, LLMs are prompted to generate ``$d_1$'' or ``$d_2$'' to answer the question on ``which of the passages is more relevant to the query?'' See Appendix~\ref{sec:prompt_preference} for the prompt. 
Based on the results and the consistency of results when switching the order of $d_1$ and $d_2$ in the prompt, we could have $d_1$ consistently better ($d_1 > d_2$), $d_2$ consistently better ($d_1 < d_2$), and inconsistent judgement ($d_1 = d_2$), as the LLM generated preferences.

To get a consistent ranking from these pairwise preferences, we follow~\citet{qin2023large} to compute a ranking score $s_i$ for each document $d_i$ by performing a global aggregation on all other candidates of the same query,
\begin{equation}
\label{eq:score}
    \hat{s}_i = 1 \times \sum_{j\neq i} \mathbb{I}_{d_i > d_j} + 0.5 \times \sum_{j\neq i} \mathbb{I}_{d_i=d_j},
\end{equation}
where $\mathbb{I}_{cond}$ is an indicator function of the condition ${cond}$: 1 when cond is true and 0 otherwise. $\hat{s}_i$ essentially counts number of wins for each document. We then sort the candidates by their ranking scores $\{\hat{s}\}_q$ to get predicted ranking $\{r\}_q$. 

The ranking performance is evaluated by the normalized discounted cumulative gain (NDCG) metric:
\begin{align}
\label{eq:ndcg}
    {\rm DCG}_q &= \sum_{i\in \{d\}_q}\frac{2^{y_i}-1}{\log_2(1+r_i)}, \\
    {\rm NDCG}_q &= \frac{1}{{\rm DCG}_q^{\rm ideal}}{\rm DCG}_q,
\end{align}
where 
${\rm DCG}_q^{\rm ideal} = \max_{\{r\}_q}{\rm DCG}_q$ is the optimal DCG obtained by sorting documents by their labels~\cite{jarvelin2002cumulated}.
In practice, the ${\rm NDCG}@k$ metric that cuts off at the top $k$ results is used.

\subsection{The Consolidation Problem}
Although the two formulations, relevance and ranking predictions, are conceptually aligned to the same ground-truth labels,
different modes above are leveraged in practice for different purposes: 
the pseudo-rater mode of LLMs, directly predicting the candidate relevance to a query, gives relatively good relevance estimation $\hat{y}$~\cite{liang2022holistic}, while the ranker mode of LLMs, using pairwise prompting, achieves significantly better NDCG but with totally uncalibrated ranking scores $\hat{s}$ that have poor relevance prediction performance~\cite{qin2023large}, or see \figref{fig:rankingrater} for an example. How to address this dichotomy then is the problem that we study in this paper.

In the optimization problem with multiple objectives like this, optimizing for both relevance prediction and ranking performance, the success is difficult to be measured with a single metric. Additionally, a tradeoff typically exists between these metrics (ECE and NDCG in our case) -- improving one leading to demoting the other, represented by a Pareto front in the figure of both metrics. Please see examples in \figref{fig:tradeoff}. 
An improvement against the baselines is qualified by whether the new method could push the Pareto front by positing metrics on the better side of the current Pareto front.
\section{The Methods} \label{sec:methods}


This section presents our post-processing methods to consolidate the ranking scores $\hat{s}$ as well as the pairwise preferences from the LLM ranker mode and the relevance estimation $\hat{y}$ from the pseudo-rater mode, aiming to optimally balance ranking and relevance prediction performance.
To make a fair comparison with previous LLM rankers, we stick to zero-shot prompting results with no training or finetuning.


Specifically, we introduce a constrained regression method to find minimal perturbations of the relevance predictions $\hat{y}$ such that the resulting ranking matches the the pairwise preference predictions of PRP. 
Additionally, we also introduce an efficient version of our constrained regression method that avoids querying an LLM to construct the complete quadratic number of pairwise constraints by selecting a linear-complexity subset of pairwise comparisons.
Finally, with the constrained regression to consolidate, we propose a ranking-aware pseudo-rater pipeline that leverages both rating and ranking capabilities of LLMs to make high-quality ratings for search.

\subsection{Constrained Regression}
\label{sec:constrainedreg}

The goal of the constrained regression methods is to adjust the LLM relevance predictions $\hat{y}$ so that their order aligns with the ranking order of the PRP results $\hat{s}$.
By minimizing the perturbations to adjust the predictions, the resulting scores should closely match the original relevance predictions while adhering to the PRP's ranking performance.

Formally, given a query $q$, we aim to find a set of minimal linear modifications $\{\delta\}_q$ of the LLM relevance predictions, so that for a PRP pairwise preference $d_i > d_j$ or $\hat{s}_i > \hat{s}_j$, the modified predictions match that order: $\hat{y}_i + \delta_i > \hat{y}_j + \delta_j$.
In general terms:
\begin{eqnarray}
\label{eq:constrainedreg}
    &\{\delta^*\}_q = {\rm argmin}_{\{\delta \}_q}\sum_{i\in \{d\}_q}\delta_i^2\\
    &s.t.\ \Delta_{ij}[(\hat{y}_i + \delta_i) - (\hat{y}_j + \delta_j)]\geq 0 \nonumber\\
    &\textrm{for} \quad\forall i,j\in \{d\}_q, \nonumber
\end{eqnarray}
where $\Delta_{ij} = \hat{s}_i - \hat{s}_j$, if preference is constructed from ranking scores, or $\Delta_{ij} = \mathbb{I}_{d_i > d_j} - \mathbb{I}_{d_i < d_j}$ if direct preference is considered.
Thus, the sign of $\Delta_{ij}$ indicates the pairwise order between $i$ and $j$, and a lack of preference in ordering results in $\Delta_{ij}=0$. 
We use $\{\hat{y}+\delta^*\}$ as our final predictions for both ranking and relevance. 

The mathematical problem posed in \eqnref{eq:constrainedreg} is a well-known constrained regression problem that can easily be solved with publicly available existing math libraries~\cite{virtanen2020scipy}.

\begin{figure}[tb]
  \centering
  \includegraphics[width=.7\columnwidth]{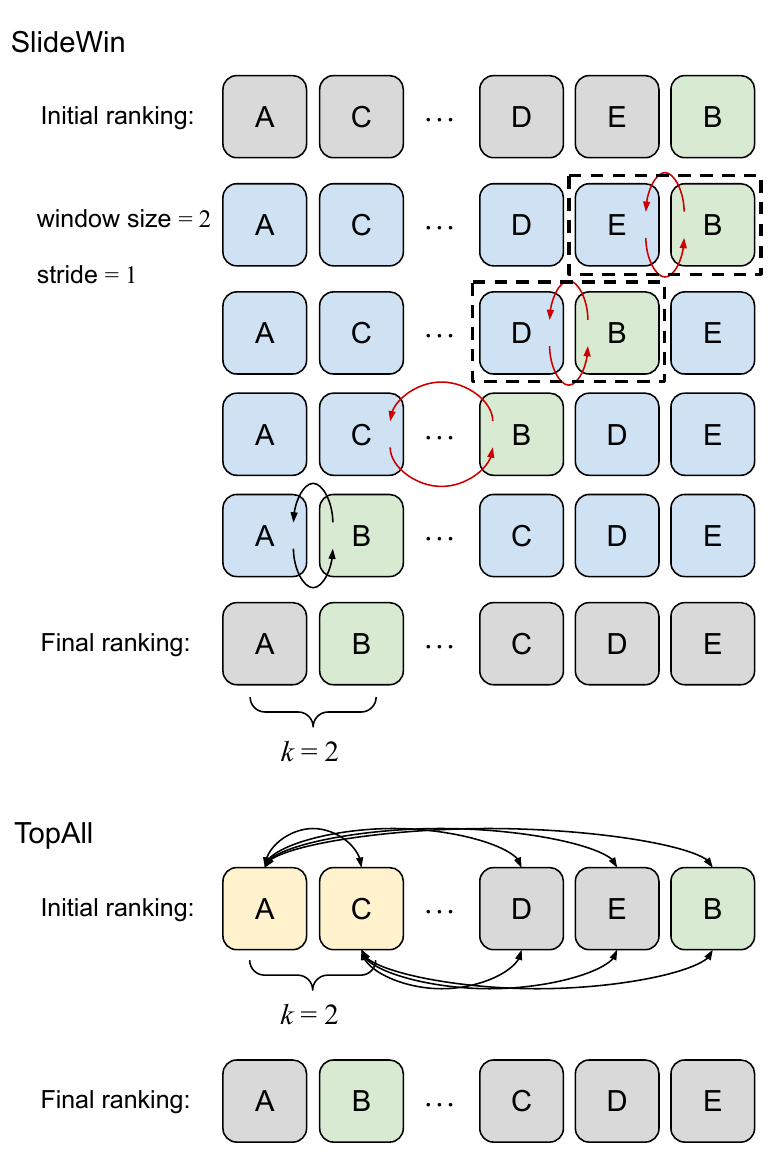}
  \caption{Illustration of how to select LLM pairwise constraints in SlideWin and TopAll methods. Top: SlideWin method with window size 2 and stride 1 takes $o(kn)$ successive pair comparisons, illustrated by paired arrows, to sort for top $k$ results from some initial ranking. Bottom: TopAll method considers top-$k$ results from an initial ranking and their pairwise constraints with all other results, shown by $o(kn)$ double-headed arrows.}
  \label{fig:methods}
\end{figure}

\subsection{Efficiency Improvements}
Constrained regression is a traditional, fast, and cost-efficient algorithm compared to LLM operations, as detailed in \secref{sec:cost}. 
A limitation of the above method is the need to identify all $o(n^2)$ pairwise constraints through pairwise ranking prompting to calculate ranking scores $\hat{s}$ in~\eqnref{eq:score} for a list of size $n$. 
As the method only depends on pairwise constraints given by $\Delta_{ij}$, a simple way to improve efficiency is to reduce the number of pair constraints to be processed by LLM. 

Here we introduce two efficient constraint choices: SlideWin and TopAll, as illustrated in \figref{fig:methods}. (1) As the ranking performance focuses mostly on the top results (top 10 or top 20), PRP work~\citep{qin2023large} proposes to just run a sliding window sorting from some initial ranking to find the top-$k$ results with $o(kn)$ pair comparisons. We just reuse these $o(kn)$ pair comparisons as constraints $\Delta_{ij}$ in \eqnref{eq:constrainedreg}. We call this variant SlideWin. (2) As our final predictions rely upon the relevance scores $\hat{y}$, we don't need to sort from random. Assuming the initial ranking from initial relevance scores $\hat{y}$ is close to the final PRP ranking, we can just consider pairwise constraints between the candidates of top relevance predictions and the rest. In specific, we consider top-$k$ in the relevance scores $\hat{y}$ and all other results in the candidate list, or top-$k$ vs.\ all, where $o(kn)$ pair constraints to be enforced. We call this variant TopAll.


\begin{table}[htbp]
\caption{Summary of constrained regression methods vs Pseudo-Rater and PRP baselines.}
\vspace{-0.2cm}
\label{tbl:method}
\resizebox{\linewidth}{!}{
\begin{tabular}{cccc}
\hline
 Methods & Use $\hat{y}$ & Use $\{d_i > d_j\}$ & \begin{tabular}{@{}c@{}}Complexity of\\LLM calls\end{tabular}  \\
\hline
 PRater  & Yes                                     & No                     & $o(n)$ \\
 PRP     & No                                      & Yes, all               & $o(n^2)$ \\
 Allpair & Yes                                     & Yes, all               & $o(n^2)$ \\
 SlideWin & Yes                                    & Yes, partial           & $o(n)$ \\
 TopAll   & Yes                                    & Yes, partial           & $o(n)$ \\
\hline
\end{tabular}
}
\end{table}

In \tblref{tbl:method}, we summarize the use of LLM-generated relevance predictions $\hat{y}$ and pairwise preferences $\{d_i > d_j\}$ and the method complexities in terms of LLM calls of all proposed methods together with the Pseudo-rater and PRP baselines.

\subsection{Ranking-Aware Pseudo-Rater}
To conclude, we propose an end-to-end ranking-aware pseudo-rater pipeline that leverages both the rating and ranking capabilities of LLMs, as illustrated in ~\figref{fig:rankingrater}.
For a given query $q$ and a list of candidate documents $\{d\}_q$, we  formulate pointwise rating and pairwise ranking prompts, then feed these prompts to the central LLM to obtain  initial ratings and pairwise preferences, respectively.
We then combine the initial ratings and pairwise preferences using our constrained regression methods for consolidation. The output of this pipeline is the ranking-aware pseudo labels.
\section{Experiment Setup}
\label{sec:exp}
We conduct experiments using several public ranking datasets to answer the following research questions:

\begin{itemize}[leftmargin=*]
\setlength\itemsep{-0.1em}
    \item {\bf RQ1}: Can our proposed constrained regression methods effectively consolidate the ranking performance of PRP and the relevance performance of LLMs as psuedo-raters?
    \item {\bf RQ2}: What is the tradeoff between ranking and relevance prediction performance for different methods?
\end{itemize}

\begin{table}[htbp]
\caption{Statistics of experimental datasets.}
\vspace{-0.2cm}
\label{tbl:dataset}
\resizebox{\linewidth}{!}{
\begin{tabular}{lccc}
\hline
 & {$\#$ of} &  & normalized \\
 {Dataset} & queries & labels & labels\\
\hline
       TREC-DL2019 & 43 & \{0, 1, 2, 3\} & \{0, 1/3, 2/3, 1\}\\
       TREC-DL2020 & 54 & \{0, 1, 2, 3\} & \{0, 1/3, 2/3, 1\} \\
       TREC-Covid  & 50 & \{0, 1, 2\} & \{0, 1/2, 1\}  \\
       DBPedia     & 400 & \{0, 1, 2\} & \{0, 1/2, 1\}  \\
       Robust04    & 249 & \{0, 1, 2\}  & \{0, 1/2, 1\}  \\
\hline
\end{tabular}
}
\end{table}

\subsection{Datasets}
We consider the public datasets with multi-level labels to study the above research questions. 
Specifically, we utilize the test sets of TREC-DL2019 and TREC-DL2020 competitions, as well as those from TREC-Covid, DBPedia, and Robust04 in the BEIR dataset~\cite{thakur2021beir}. \tblref{tbl:dataset} summarizes the statistics of queries and the range of labels. The candidate documents are selected from the MS MARCO v1 passage corpus, which contains 8.8 million passages. LLM rankers are applied on the top 100 passages retrieved by BM25~\cite{Lin_etal_SIGIR2021_Pyserini} for each query, same setting as existing LLM ranking works~\cite{baidugpt,jimmygpt, qin2023large}.

\begin{table*}[tb]
\caption{Evaluation of LLM-based ranking methods on both ranking (NDCG@10) and relevance prediction (ECE and MSE) metrics on TREC-DL 2019 and 2020, TREC-Covid, DBPedia, and Robust04. Bold numbers are the best of all and numbers underlined are the best among proposed methods in each row. Upscript ``$\dagger$'' indicate statistical significance with p-value=0.01 of better performance than the baselines, PRater for NDCG@10 and PRP for ECE and MSE.}
\vspace{-0.2cm}
\label{tbl:result}
\centering
\resizebox{1.\linewidth}{!}{
\begin{tabular}{llcccccccc}
\hline
&& \multicolumn{5}{c}{Baselines} & \multicolumn{3}{c}{Our Consolidation Methods} \\
& Method &  \multicolumn{1}{|c}{BM25}  & PRater &  PRP  &  \multicolumn{1}{|c}{PRater+PWL}  & PRP+PWL & \multicolumn{1}{|c}{Allpair} &    SlideWin &    {TopAll} \\
\hline
\hline
\multirow{2}{*}{TREC-DL2019} & NDCG@10 & 0.5058 & 0.6461 & 0.7242 & 0.6461 & 0.7242 & 0.7236$^\dagger$ & \underline{\bf 0.7265}$^\dagger$ & 0.7189$^\dagger$ \\
                             & ECE     & 0.2088 & 0.1167 & 0.3448 & 0.1199 & 0.1588 & \underline{\bf 0.1084}$^\dagger$ & 0.1090$^\dagger$ & 0.1199$^\dagger$ \\
                             & MSE     & 0.1096 & 0.0688 & 0.1787 & 0.0652 & 0.0836 & \underline{\bf 0.0592}$^\dagger$ & 0.0601$^\dagger$ & 0.0692$^\dagger$ \\
\hline
\multirow{2}{*}{TREC-DL2020} & NDCG@10 & 0.4796 & 0.6539 & {\bf 0.7069} & 0.6539 & {\bf 0.7069} & \underline{0.7054}$^\dagger$ & 0.7046$^\dagger$ & 0.7025 \\
                             & ECE     & 0.2219 & 0.0991 & 0.3690 & {\bf 0.0793} & 0.0954 & \underline{ 0.0865}$^\dagger$ & 0.0911$^\dagger$ & 0.0966$^\dagger$ \\
                             & MSE     & 0.1122 & 0.0632 & 0.1978 & {\bf 0.0444} & 0.0488 & \underline{0.0519}$^\dagger$ & 0.0560$^\dagger$ & 0.0600$^\dagger$ \\
\hline
\multirow{2}{*}{TREC-Covid}       & NDCG@10 & 0.5947 & 0.7029 & {\bf 0.8231} & 0.7029 & {\bf 0.8231} & \underline{0.8220}$^\dagger$ & 0.7943$^\dagger$ & 0.7962$^\dagger$ \\
                             & ECE     & 0.2460 & 0.2047 & 0.2340  & {\bf 0.1590} & 0.2192 & 0.1990$^\dagger$ & \underline{0.1984}$^\dagger$ & 0.2216 \\
                             & MSE     & 0.2268 & 0.1756 & 0.1621  & {\bf 0.1419} & 0.1557 & \underline{0.1575}$^\dagger$ & 0.1644 & 0.1870 \\
\hline
\multirow{2}{*}{DBPedia}     & NDCG@10 & 0.3180 & 0.3057 & 0.4613 & 0.3057 & 0.4613 & 0.4598$^\dagger$ & \underline{\bf 0.4651}$^\dagger$ & 0.4029$^\dagger$ \\
                             & ECE     & 0.2183 & 0.1360 & 0.4364 & {\bf 0.0554} & 0.0629 & \underline{ 0.1302}$^\dagger$ & 0.1308$^\dagger$ & 0.1329$^\dagger$ \\
                             & MSE     & 0.0864 & 0.0967 & 0.2571 & 0.0387 & {\bf 0.0350} & \underline{0.0846}$^\dagger$ & 0.0863$^\dagger$ & 0.0901$^\dagger$ \\
\hline
\multirow{2}{*}{Robust04}      & NDCG@10  & 0.4070 & 0.5296 & {\bf 0.5551} & 0.5296 & {\bf 0.5551} & \underline{0.5532}$^\dagger$ & 0.5364 & 0.5347 \\
                             & ECE      & 0.1291 & {\bf 0.0650} & 0.4154 & 0.0689 & {0.0658} & \underline{0.0654}$^\dagger$ & 0.0669$^\dagger$ & 0.0804$^\dagger$ \\
                             & MSE      & 0.0594 & 0.0386 & 0.2285 & 0.0368 & {\bf 0.0361} & \underline{0.0379}$^\dagger$ & 0.0390$^\dagger$ & 0.0509$^\dagger$ \\
\hline
\end{tabular}
}
\end{table*}


\subsection{Evaluation Metrics}
For  ranking performance, we adopt NDCG (as defined in \eqnref{eq:ndcg}) as the evaluation metric, with higher values indicating better performance.
We primarily focus on NDCG@10, but also present NDCG with other cutoff points in certain ablation studies. 
For the relevance prediction performance, we use the {\it mean squared error} (MSE) in \eqnref{eq:mse} and the {\it empirical calibration error} (ECE) in \eqnref{eq:ece} as the evaluation metrics. 
The lower ECE values indicate better relevance predictions. 
In this work, we choose $M=10$ bins~\citep{naeini2015obtaining} with each bin containing approximately the same number  of documents ($\sim10$ documents per bin). 


\subsection{Comparison Methods}
We investigate the  performance of the following methods in ranking and relevance prediction:
\begin{itemize}[leftmargin=*]
    \item \textbf{BM25}~\cite{Lin_etal_SIGIR2021_Pyserini}: The sole non-LLM ranker baseline.
    \item \textbf{PRater}~\cite{baidugpt}: The pointwise LLM relevance pseudo-rater approach.
    \item \textbf{PRP}~\cite{qin2023large}: The LLM ranker using pairwise ranking prompting (PRP). All pair comparisons are used to compute the ranking scores (as in~\eqnref{eq:score}).
    \item \textbf{Allpair} (Ours): The naive constrained regression method in \eqnref{eq:constrainedreg} with all pairwise preferences based on the PRP scores, $\Delta_{ij}=\hat{s}_i-\hat{s}_j$.
    \item \textbf{SlideWin} (Ours): The constrained regression method in \eqnref{eq:constrainedreg} with pairwise LLM constraints collected with the sliding window ordering approach, proposed by \citet{qin2023large}: pair comparisons are selected from sliding bottom up on the initial order by BM25 scores with sliding window size $k=10$. 
    \item \textbf{TopAll} (Ours): The constrained regression method with pairwise LLM constraints on the pairs between top $k=10$ results by sorting on pseudo-rater predictions $\hat{y}$ versus all candidates in the list.
\end{itemize}
Unless specified, all LLM results in above methods are based on the FLAN-UL2 model~\cite{ul2}, an OSS LLM~\footnote{https://huggingface.co/google/flan-ul2}.

In addition, motivated by the multi-objective approach to consolidate ranking and relevance predictions in non-LLM rankers~\cite{yan2022scale}, we also consider a simple weighted ensemble of PRater predictions $\hat{y}$ and PRP scores $\hat{s}$:
\begin{equation}
\label{eq:ensemble}
    \hat{y} + w \hat{s},
\end{equation}
where $w$ is the relative weight, and we use \textbf{Ensemble} to refer the method. Note that in practice some labeled data is needed to decide $w$, while the other methods discussed above are fully unsupervised.

\subsection{Prediction Normalization}
It should be noted that none of the  methods are optimized for ground truth label values, hence, the ECE and MSE metrics from the raw results are not directly comparable.
Thus, we scale their predictions to match the range of the ground truth labels:
\begin{equation}
\label{eq:rescale}
    \tilde{y} = y_{\rm min} + (y_{\rm max} - y_{\rm min}) \frac{\hat{y} - \min(\hat{y})}{\max(\hat{y}) - \min(\hat{y})},
\end{equation}
where $\max$ and $\min$ are global max and global min on the full test set.
Subsequently, we compute ECE based on the scaled predicted scores $\tilde{y}$.
For normalized relevance labels, we insert $y_{\rm min}=0$ and $y_{\rm max}=1$.

\subsection{Supervised PWL Transformation}
We also compare a post-processing method \textit{requiring labelled data},
specifically the \emph{piecewise linear transformation} (PWL) introduced in \citet{ravina2021distilling}, defined as follows,
\begin{align}
\label{eq:pwl}
    &f\big(s \mid \{\tilde{s}_m, \tilde{y}_m\}_{m=1}^M\big) = \\
    &\quad\begin{cases}
        \tilde{y}_1 & s\leq \tilde{s}_1, \\
        \tilde{y}_m + \frac{\tilde{y}_{m+1} - \tilde{y}_m}{\tilde{s}_{m+1} - \tilde{s}_m}(s - \tilde{s}_m)
        & \tilde{s}_m < s\leq \tilde{s}_{m+1}, \\
        \tilde{y}_M & s> \tilde{s}_M, 
    \end{cases}
    \nonumber
\end{align}
where $\{\tilde{s}_m, \tilde{y}_m\}_{m=1}^M$ are $2M$ fitting parameters. $\tilde{y}_{m+1} > \tilde{y}_m$ and $\tilde{s}_{m+1} > \tilde{s}_m$ are enforced for any $m$ to reinforce the monotonicity of the transformation to effectively scales predictions without affecting the ranking order.

We apply PWL to baseline methods PRater and PRP as a special set of baselines with labelled data available, named as \textbf{PRater+PWL} and \textbf{PRP+PWL} in the results. Comparing these with supervised methods allow for a better understanding of our proposed unsupervised approaches. To compute the post-fitting in PWL, we apply four-fold cross-validation to the test set data: we randomly divide the test set into four folds by queries, and then fit the PWL transformation function on one set and predict on one of the others, repeatedly, to get PWL transformation results for the whole test set.

\begin{figure}[tb]
  \includegraphics[width=.451\columnwidth]{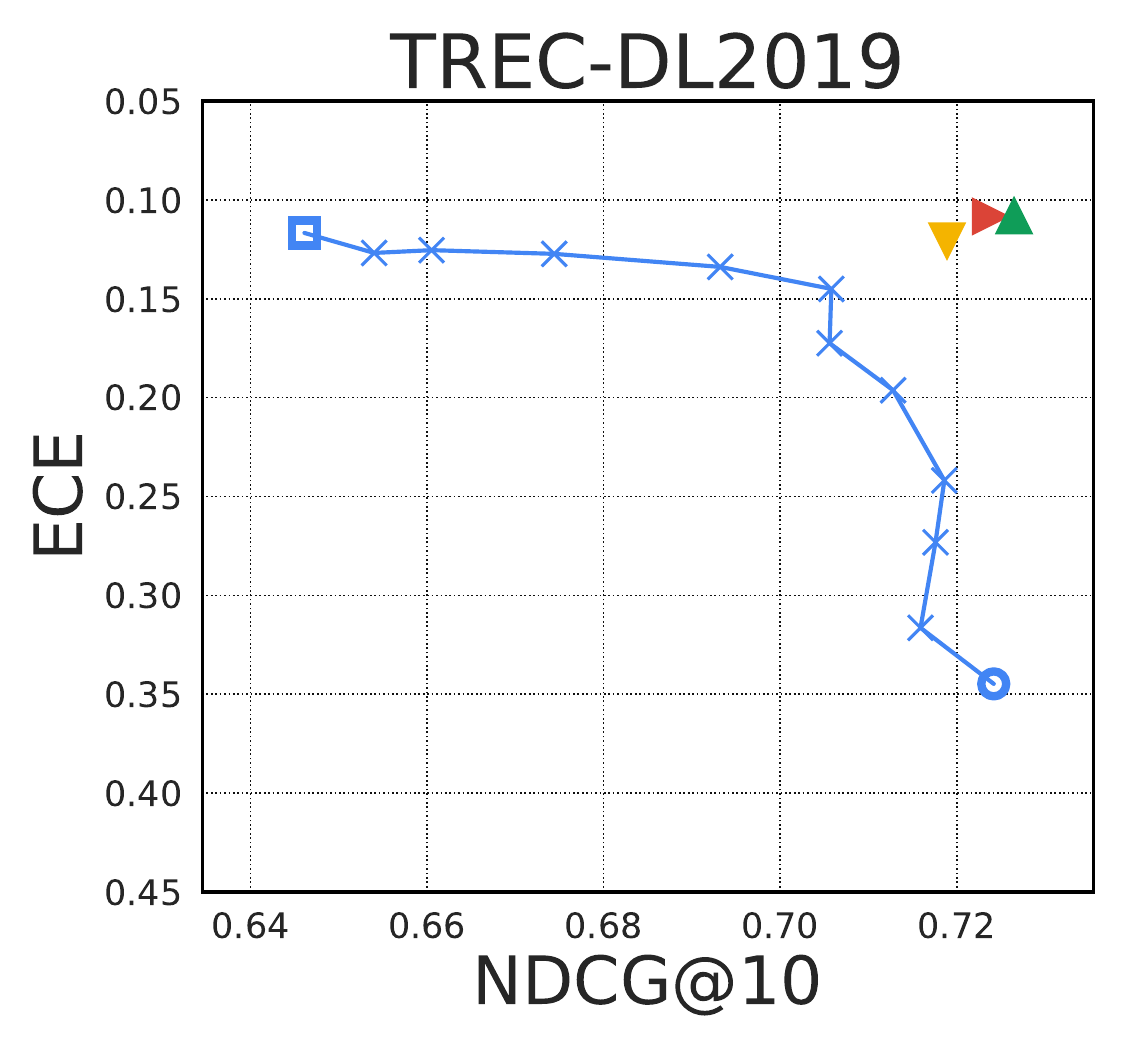}
  \includegraphics[width=.42\columnwidth]{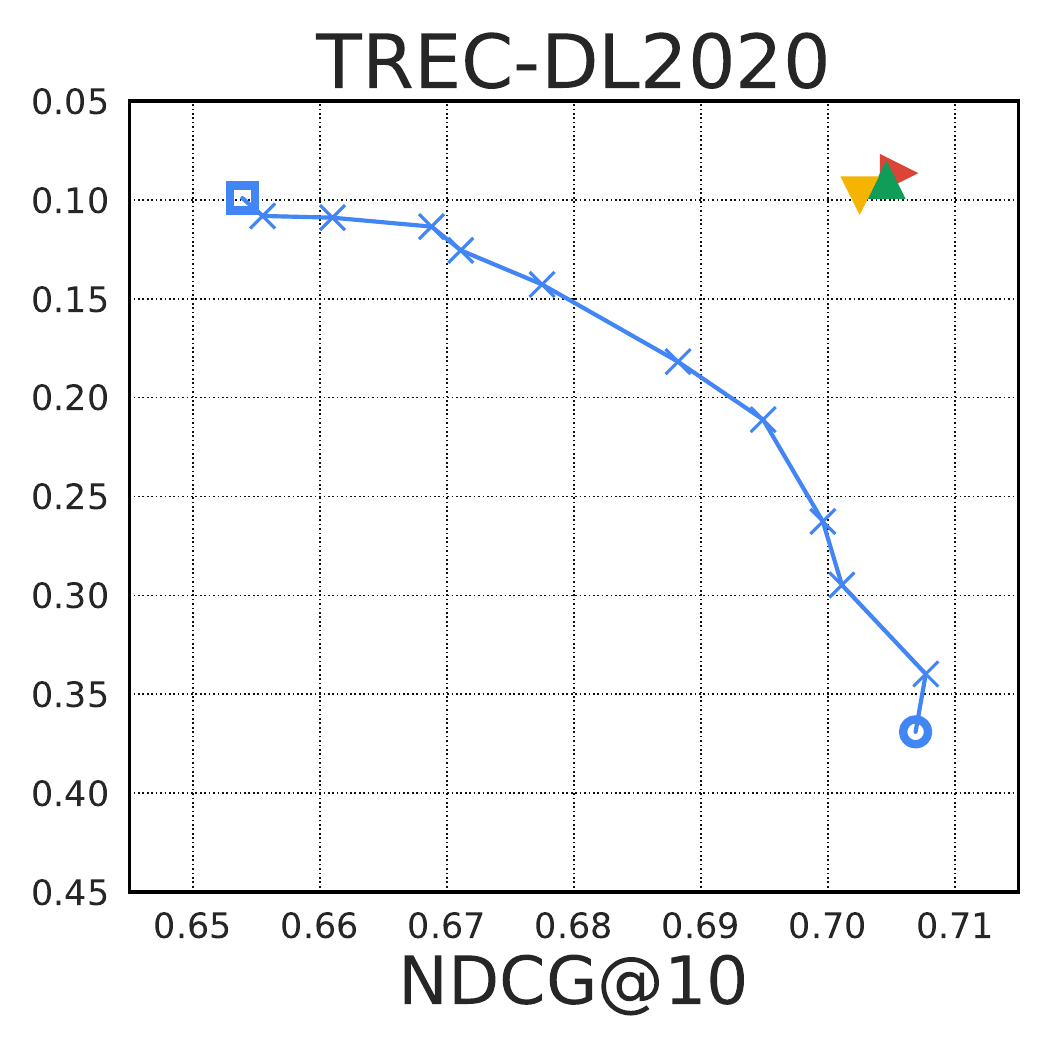}
  \includegraphics[width=.451\columnwidth]{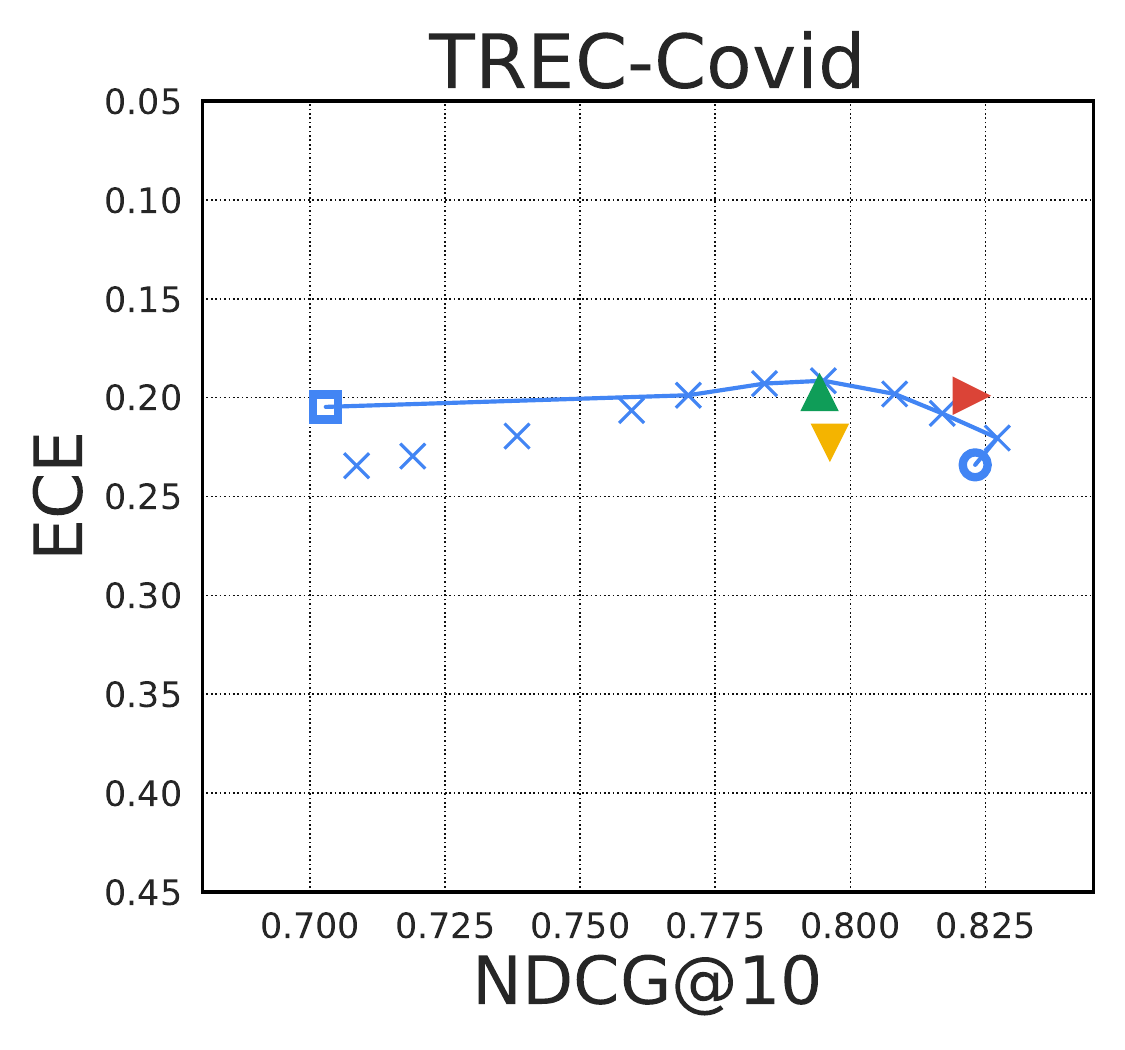}
  \includegraphics[width=.42\columnwidth]{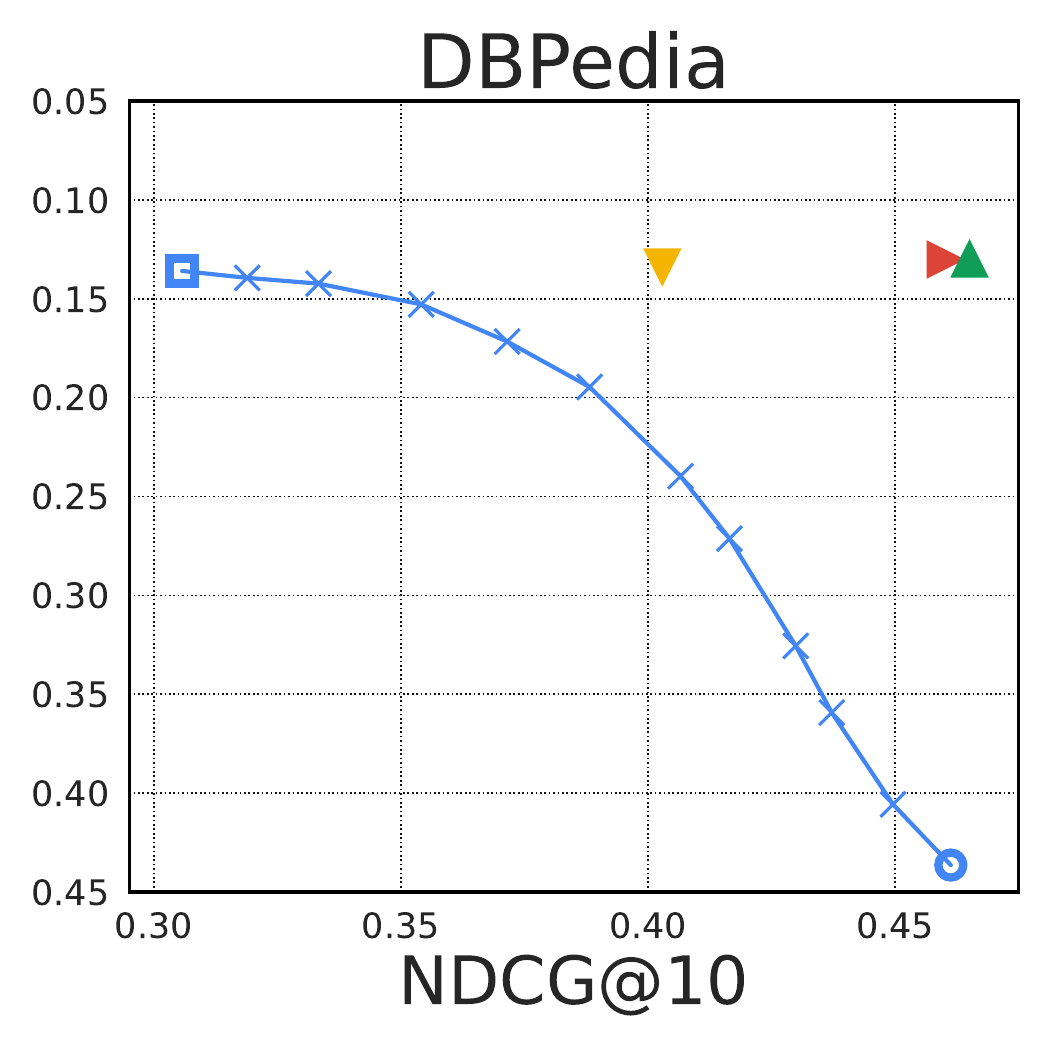}
  \includegraphics[width=.82\columnwidth]{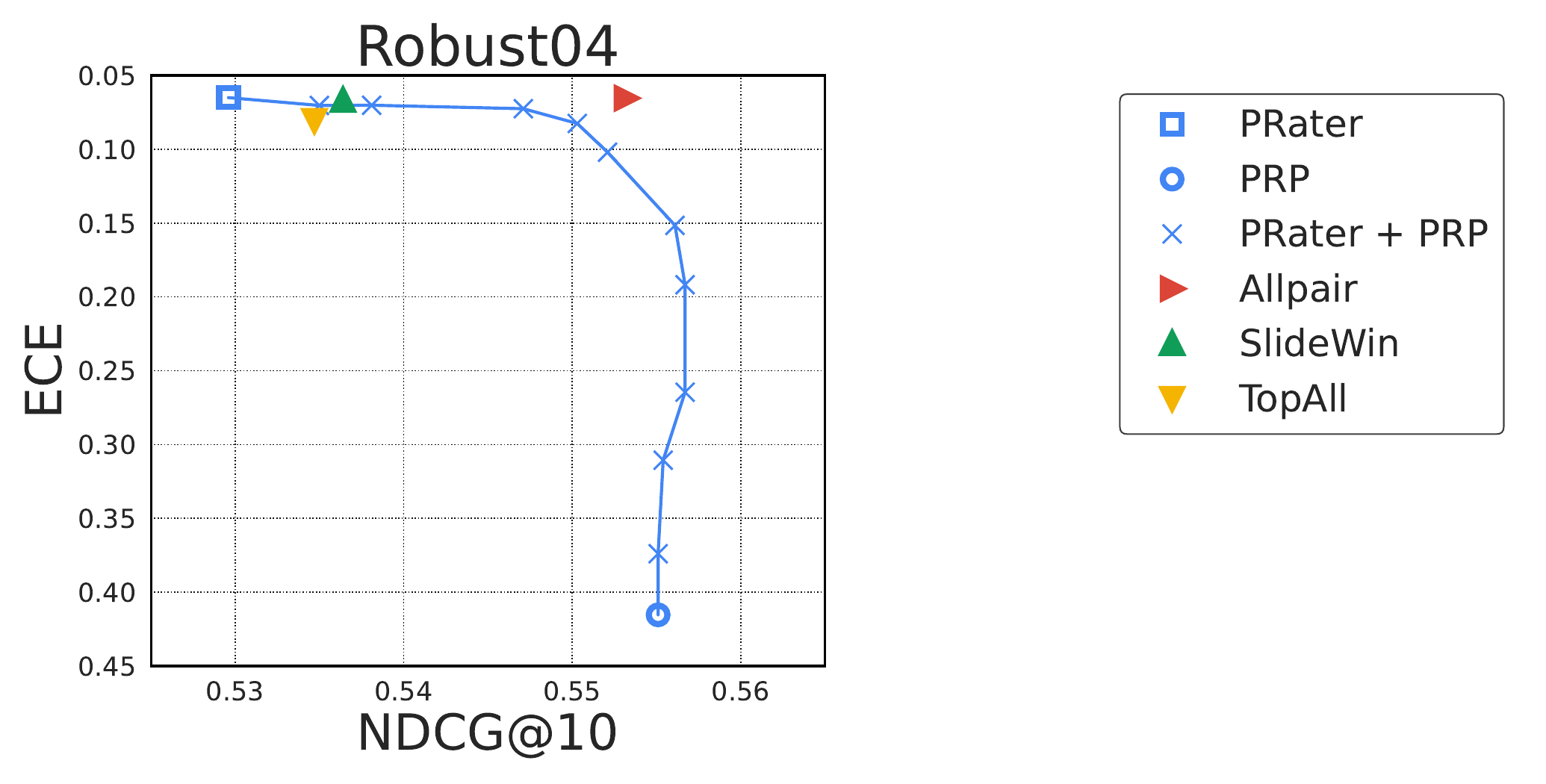}
  \caption{{Tradeoff plots} on ECE versus NDCG@10 on five ranking datasets. NDCG@10 is higher the better and ECE is lower the better. Overall better methods are on the top right corner of the plots. Lines correspond to the Pareto fronts of Ensemble of PRater and PRP by tuning the weight $w$ in \eqnref{eq:ensemble}. Our consolidation methods in \tblref{tbl:result} are scattered in the Figure.}
  \label{fig:tradeoff}
\end{figure}

\section{Experimental Results}
\label{sec:results}

\subsection{Main Results}
The main results, summarized in~\tblref{tbl:result} and \figref{fig:tradeoff}, include the following observations: 
\begin{itemize}[leftmargin=*]
    \item MSE and ECE metrics are consistent in \tblref{tbl:result}. Therefore, we will focus on ECE for the remainder of the discussion.
    \item Without PWL transformations, the pointwise relevance LLM rater (PRater) performs better in labelling than both the naive BM25 and PRP rankers, as evidenced by a consistenly lower ECE in~\tblref{tbl:result}.
    \item PRater is also better in the sense of label distribution: after PWL transformation, PRater has a lower ECE than PRP. 
    \item Despite its poor ECE, PRP has the best or nearly best ranking performance in terms of NDCG.
    \item Constrained regression approach can best leverage the relevance estimations of PRater and the ranking capability of PRP and reaches comparable ranking performance in terms of NDCG to PRP, and on par or even better relevance prediction in terms of ECE to PRater.
    \item Our methods consolidate the ranking from PRP and relevance predictions from PRater effectively, evident by that the combined performance on NDCG and ECE sits well beyond the Pareto fronts of simple weighted Ensemble of the two.
    \item Our consolidation methods even outperform PRP+PWL, the one with extra data, in ECE on 4 out of 5 datasets and while keep ranking performance in NDCG@10 as good on all datasets. This is because supervised methods may not learn effectively with limited annotations, which is the case for public search datasets given the high cost of collecting human annotations.
    \item Finally, efficient constrained regression methods may trade off some performance in ranking and regression for the efficiency, but they can still outperform the baselines of PRater and PRP and weighted ensemble of the two in most of the datasets.
\end{itemize}

With these main results, we could answer the main research questions. 
{\bf RQ1.} Using the constrained regression methods, we could boost the LLM raters in the superior ranking capability of PRP rankers while keep their relevance predictions nearly untouched. 
{\bf RQ2.} Naive ensemble of LLM pseudo-rater predictions and PRP scores may lead to a tradeoff between ranking and relevance prediction performance. However, we could get over this tradeoff with the constrained regression methods.

\begin{table}[htbp]
\caption{Model size effect of constrained regression methods and LLM baselines on TREC-DL 2020.}
\vspace{-0.2cm}
\label{tbl:result_size}
\resizebox{\linewidth}{!}{
\begin{tabular}{ccccc}
\hline
& \multicolumn{2}{c}{NDCG@10} & \multicolumn{2}{c}{ECE} \\
{Method} & T5-XXL & \multicolumn{1}{c}{UL2} & T5-XXL & UL2 \\
\hline
\hline 
{PRater}  & 0.6258 & \textbf{0.6539} & \textbf{0.0949} & 0.0991 \\
{PRP}    & 0.6985 & \textbf{0.7069} & 0.3698 & \textbf{0.3690} \\
{Allpair} & 0.6960 & \textbf{0.7054} & 0.0871 & \textbf{0.0865} \\
{SlideWin} & 0.6735 & \textbf{0.7046} & \textbf{0.0900} & 0.0911 \\
{TopAll} & 0.6794 & \textbf{0.7025}  &  0.1038 & \textbf{0.0966} \\
\hline
\end{tabular}
}
\end{table}

\subsection{Model Size Effect}

As with other tasks involving pretrained LLMs, larger models generally perform better in both ranking and regression metrics. \tblref{tbl:result_size} shows that our constrained regression methods achieve significantly better NDCG, and comparable or better ECE with the FLAN-UL2 model compared to the FLAN-T5-XXL model~\footnote{https://huggingface.co/google/flan-t5-xxl}. The same size effect is observed in PRater and PRP as well. This shows our consolidation method scales together with the underlying LLM's performance. 

We have also run experiments on the choices of initial ranking models and choices of parameter $k$ for efficient constrained regression methods (SlideWin and TopAll). The results are included in \secref{sec:more}.

\section{Conclusion} \label{sec:conclu}
In this work, we have studied the problem of consolidating ranking and relevance predictions of LLMs. We have found that the direct scores from the zero-shot pairwise ranking prompting (PRP) poorly correlate with ground truth labels. To leverage the superior ranking ability of PRP while aligning closely with the ground truth labels, we have investigated post-processing methods and proposed a class of constrained regression methods that combine pointwise ratings from the LLM raters and pairwise constraints from the PRP rankers to take advantage of the two.
We have demonstrated with experiments on public ranking datasets that our methods are efficient and effective, offering competitive or superior ranking performance compared to the PRP baseline and relevance prediction performance akin to the pointwise LLM rater.
Last but not least, we have proposed a novel framework on how to effectively use generative LLMs to generate ranking-aware ratings, foundation for LLM-powered search ranking.

\section*{Limitations}
\label{sec:limitations}
First, our work mainly focused on consolidating relevance raters with pairwise LLM rankers due to their effectiveness, particularly with moderate-sized open-sourced LLMs. Our methods can be applied to listwise ranking results from listwise LLM rankers~\cite{sun2023chatgpt} by decomposing their ranking results into pairwise comparisons. However, more effective methods to consolidate listwise rankers, may exist, which we consider for future work.  Second, our framework assumes reasonable rating  and ranking performance by LLMs.
Although generally supported by advances in LLM research and validated across diverse datasets, more advanced adjustments may be required for scenarios where LLMs perform suboptimally, such as in domains opaque to the underlying LLMs. 

\bibliography{references}

\begin{thebibliography}{36}
\expandafter\ifx\csname natexlab\endcsname\relax\def\natexlab#1{#1}\fi

\bibitem[{Bai et~al.(2023)Bai, Jagerman, Qin, Yan, Kar, Lin, Wang, Bendersky,
  and Najork}]{bai2023regression}
Aijun Bai, Rolf Jagerman, Zhen Qin, Le~Yan, Pratyush Kar, Bing-Rong Lin,
  Xuanhui Wang, Michael Bendersky, and Marc Najork. 2023.
\newblock Regression compatible listwise objectives for calibrated ranking with
  binary relevance.
\newblock In \emph{Proceedings of the 32nd ACM International Conference on
  Information and Knowledge Management}, pages 4502--4508.

\bibitem[{Bonifacio et~al.(2022)Bonifacio, Abonizio, Fadaee, and
  Nogueira}]{bonifacio2022inpars}
Luiz Bonifacio, Hugo Abonizio, Marzieh Fadaee, and Rodrigo Nogueira. 2022.
\newblock {InPars}: Unsupervised dataset generation for information retrieval.
\newblock In \emph{Proceedings of the 45th International ACM SIGIR Conference
  on Research and Development in Information Retrieval}, pages 2387--2392.

\bibitem[{Google et~al.(2023)Google, Anil, Dai, Firat, Johnson, Lepikhin,
  Passos, Shakeri, Taropa, Bailey, Chen, Chu, Clark, Shafey, Huang,
  Meier-Hellstern, Mishra, Moreira, Omernick, Robinson, Ruder, Tay, Xiao, Xu,
  Zhang, Abrego, Ahn, Austin, Barham, Botha, Bradbury, Brahma, Brooks, Catasta,
  Cheng, Cherry, Choquette-Choo, Chowdhery, Crepy, Dave, Dehghani, Dev, Devlin,
  Díaz, Du, Dyer, Feinberg, Feng, Fienber, Freitag, Garcia, Gehrmann,
  Gonzalez, Gur-Ari, Hand, Hashemi, Hou, Howland, Hu, Hui, Hurwitz, Isard,
  Ittycheriah, Jagielski, Jia, Kenealy, Krikun, Kudugunta, Lan, Lee, Lee, Li,
  Li, Li, Li, Li, Lim, Lin, Liu, Liu, Maggioni, Mahendru, Maynez, Misra,
  Moussalem, Nado, Nham, Ni, Nystrom, Parrish, Pellat, Polacek, Polozov, Pope,
  Qiao, Reif, Richter, Riley, Ros, Roy, Saeta, Samuel, Shelby, Slone, Smilkov,
  So, Sohn, Tokumine, Valter, Vasudevan, Vodrahalli, Wang, Wang, Wang, Wang,
  Wieting, Wu, Xu, Xu, Xue, Yin, Yu, Zhang, Zheng, Zheng, Zhou, Zhou, Petrov,
  and Wu}]{palm2}
Google, Rohan Anil, Andrew~M. Dai, Orhan Firat, Melvin Johnson, Dmitry
  Lepikhin, Alexandre Passos, Siamak Shakeri, Emanuel Taropa, Paige Bailey,
  Zhifeng Chen, Eric Chu, Jonathan~H. Clark, Laurent~El Shafey, Yanping Huang,
  Kathy Meier-Hellstern, Gaurav Mishra, Erica Moreira, Mark Omernick, Kevin
  Robinson, Sebastian Ruder, Yi~Tay, Kefan Xiao, Yuanzhong Xu, Yujing Zhang,
  Gustavo~Hernandez Abrego, Junwhan Ahn, Jacob Austin, Paul Barham, Jan Botha,
  James Bradbury, Siddhartha Brahma, Kevin Brooks, Michele Catasta, Yong Cheng,
  Colin Cherry, Christopher~A. Choquette-Choo, Aakanksha Chowdhery, Clément
  Crepy, Shachi Dave, Mostafa Dehghani, Sunipa Dev, Jacob Devlin, Mark Díaz,
  Nan Du, Ethan Dyer, Vlad Feinberg, Fangxiaoyu Feng, Vlad Fienber, Markus
  Freitag, Xavier Garcia, Sebastian Gehrmann, Lucas Gonzalez, Guy Gur-Ari,
  Steven Hand, Hadi Hashemi, Le~Hou, Joshua Howland, Andrea Hu, Jeffrey Hui,
  Jeremy Hurwitz, Michael Isard, Abe Ittycheriah, Matthew Jagielski, Wenhao
  Jia, Kathleen Kenealy, Maxim Krikun, Sneha Kudugunta, Chang Lan, Katherine
  Lee, Benjamin Lee, Eric Li, Music Li, Wei Li, YaGuang Li, Jian Li, Hyeontaek
  Lim, Hanzhao Lin, Zhongtao Liu, Frederick Liu, Marcello Maggioni, Aroma
  Mahendru, Joshua Maynez, Vedant Misra, Maysam Moussalem, Zachary Nado, John
  Nham, Eric Ni, Andrew Nystrom, Alicia Parrish, Marie Pellat, Martin Polacek,
  Alex Polozov, Reiner Pope, Siyuan Qiao, Emily Reif, Bryan Richter, Parker
  Riley, Alex~Castro Ros, Aurko Roy, Brennan Saeta, Rajkumar Samuel, Renee
  Shelby, Ambrose Slone, Daniel Smilkov, David~R. So, Daniel Sohn, Simon
  Tokumine, Dasha Valter, Vijay Vasudevan, Kiran Vodrahalli, Xuezhi Wang,
  Pidong Wang, Zirui Wang, Tao Wang, John Wieting, Yuhuai Wu, Kelvin Xu, Yunhan
  Xu, Linting Xue, Pengcheng Yin, Jiahui Yu, Qiao Zhang, Steven Zheng,
  Ce~Zheng, Weikang Zhou, Denny Zhou, Slav Petrov, and Yonghui Wu. 2023.
\newblock \href {http://arxiv.org/abs/2305.10403} {{PaLM 2} technical report}.

\bibitem[{Guo et~al.(2017)Guo, Pleiss, Sun, and
  Weinberger}]{guo2017calibration}
Chuan Guo, Geoff Pleiss, Yu~Sun, and Kilian~Q Weinberger. 2017.
\newblock On calibration of modern neural networks.
\newblock In \emph{International conference on machine learning}, pages
  1321--1330. PMLR.

\bibitem[{Jagerman et~al.(2023)Jagerman, Zhuang, Qin, Wang, and
  Bendersky}]{jagerman2023query}
Rolf Jagerman, Honglei Zhuang, Zhen Qin, Xuanhui Wang, and Michael Bendersky.
  2023.
\newblock Query expansion by prompting large language models.
\newblock \emph{arXiv preprint arXiv:2305.03653}.

\bibitem[{J{\"a}rvelin and Kek{\"a}l{\"a}inen(2002)}]{jarvelin2002cumulated}
Kalervo J{\"a}rvelin and Jaana Kek{\"a}l{\"a}inen. 2002.
\newblock Cumulated gain-based evaluation of {IR} techniques.
\newblock \emph{ACM Transactions on Information Systems}, 20(4):422--446.

\bibitem[{Liang et~al.(2022)Liang, Bommasani, Lee, Tsipras, Soylu, Yasunaga,
  Zhang, Narayanan, Wu, Kumar et~al.}]{liang2022holistic}
Percy Liang, Rishi Bommasani, Tony Lee, Dimitris Tsipras, Dilara Soylu,
  Michihiro Yasunaga, Yian Zhang, Deepak Narayanan, Yuhuai Wu, Ananya Kumar,
  et~al. 2022.
\newblock Holistic evaluation of language models.
\newblock \emph{arXiv preprint arXiv:2211.09110}.

\bibitem[{Lin et~al.(2021)Lin, Ma, Lin, Yang, Pradeep, and
  Nogueira}]{Lin_etal_SIGIR2021_Pyserini}
Jimmy Lin, Xueguang Ma, Sheng-Chieh Lin, Jheng-Hong Yang, Ronak Pradeep, and
  Rodrigo Nogueira. 2021.
\newblock {Pyserini}: A {Python} toolkit for reproducible information retrieval
  research with sparse and dense representations.
\newblock In \emph{Proceedings of the 44th Annual International ACM SIGIR
  Conference on Research and Development in Information Retrieval (SIGIR
  2021)}, pages 2356--2362.

\bibitem[{Liu(2009)}]{8186875}
Tie-Yan Liu. 2009.
\newblock Learning to rank for information retrieval.
\newblock \emph{Foundation and Trends\textsuperscript{\textregistered} in
  Information Retrieval}, 3(3):225--331.

\bibitem[{Ma et~al.(2023)Ma, Zhang, Pradeep, and Lin}]{jimmygpt}
Xueguang Ma, Xinyu Zhang, Ronak Pradeep, and Jimmy Lin. 2023.
\newblock Zero-shot listwise document reranking with a large language model.
\newblock \emph{arXiv preprint arXiv:2305.02156}.

\bibitem[{Menon et~al.(2012)Menon, Jiang, Vembu, Elkan, and
  Ohno-Machado}]{Menon:ICML12}
Aditya~Krishna Menon, Xiaoqian Jiang, Shankar Vembu, Charles Elkan, and Lucila
  Ohno-Machado. 2012.
\newblock Predicting accurate probabilities with a ranking loss.
\newblock In \emph{Proceedings of the 29th International Conference on Machine
  Learning}, pages 703--710.

\bibitem[{Naeini et~al.(2015)Naeini, Cooper, and
  Hauskrecht}]{naeini2015obtaining}
Mahdi~Pakdaman Naeini, Gregory Cooper, and Milos Hauskrecht. 2015.
\newblock Obtaining well calibrated probabilities using bayesian binning.
\newblock In \emph{Twenty-Ninth AAAI Conference on Artificial Intelligence}.

\bibitem[{Nogueira and Cho(2019)}]{nogueira2019passage}
Rodrigo Nogueira and Kyunghyun Cho. 2019.
\newblock Passage re-ranking with {BERT}.
\newblock \emph{arXiv preprint arXiv:1901.04085}.

\bibitem[{Nogueira et~al.(2020)Nogueira, Jiang, Pradeep, and Lin}]{monot5}
Rodrigo Nogueira, Zhiying Jiang, Ronak Pradeep, and Jimmy Lin. 2020.
\newblock Document ranking with a pretrained sequence-to-sequence model.
\newblock In \emph{Findings of the Association for Computational Linguistics:
  EMNLP 2020}, pages 708--718.

\bibitem[{OpenAI(2023)}]{gpt4}
OpenAI. 2023.
\newblock {GPT-4} technical report.
\newblock \emph{arXiv preprint arXiv:2303.08774}.

\bibitem[{Platt(2000)}]{PlattScaling:1999}
John Platt. 2000.
\newblock Probabilistic outputs for support vector machines and comparisons to
  regularized likelihood methods.
\newblock In Alexander~J. Smola, Peter Bartlett, Bernhard Schölkopf, and Dale
  Schuurmans, editors, \emph{Advances in Large Margin Classifiers}, page
  61–74. MIT Press.

\bibitem[{Pradeep et~al.(2023)Pradeep, Sharifymoghaddam, and
  Lin}]{pradeep2023rankvicuna}
Ronak Pradeep, Sahel Sharifymoghaddam, and Jimmy Lin. 2023.
\newblock Rankvicuna: Zero-shot listwise document reranking with open-source
  large language models.
\newblock \emph{arXiv preprint arXiv:2309.15088}.

\bibitem[{Qin et~al.(2023)Qin, Jagerman, Hui, Zhuang, Wu, Shen, Liu, Liu,
  Metzler, Wang et~al.}]{qin2023large}
Zhen Qin, Rolf Jagerman, Kai Hui, Honglei Zhuang, Junru Wu, Jiaming Shen,
  Tianqi Liu, Jialu Liu, Donald Metzler, Xuanhui Wang, et~al. 2023.
\newblock Large language models are effective text rankers with pairwise
  ranking prompting.
\newblock \emph{arXiv preprint arXiv:2306.17563}.

\bibitem[{Qin et~al.(2021)Qin, Yan, Zhuang, Tay, Pasumarthi, Wang, Bendersky,
  and Najork}]{dasalc}
Zhen Qin, Le~Yan, Honglei Zhuang, Yi~Tay, Rama~Kumar Pasumarthi, Xuanhui Wang,
  Michael Bendersky, and Marc Najork. 2021.
\newblock Are neural rankers still outperformed by gradient boosted decision
  trees?
\newblock In \emph{Proceedings of the 9th International Conference on Learning
  Representations}.

\bibitem[{Ravina et~al.(2021)Ravina, Sterling, Oryeshko, Bell, Zhuang, Wang,
  Wu, and Grushetsky}]{ravina2021distilling}
Walker Ravina, Ethan Sterling, Olexiy Oryeshko, Nathan Bell, Honglei Zhuang,
  Xuanhui Wang, Yonghui Wu, and Alexander Grushetsky. 2021.
\newblock Distilling interpretable models into human-readable code.
\newblock \emph{arXiv preprint arXiv:2101.08393}.

\bibitem[{Sachan et~al.(2022)Sachan, Lewis, Joshi, Aghajanyan, Yih, Pineau, and
  Zettlemoyer}]{sachan2022improving}
Devendra~Singh Sachan, Mike Lewis, Mandar Joshi, Armen Aghajanyan, Wen-tau Yih,
  Joelle Pineau, and Luke Zettlemoyer. 2022.
\newblock Improving passage retrieval with zero-shot question generation.
\newblock \emph{arXiv preprint arXiv:2204.07496}.

\bibitem[{Sun et~al.(2023{\natexlab{a}})Sun, Yan, Ma, Ren, Yin, and
  Ren}]{baidugpt}
Weiwei Sun, Lingyong Yan, Xinyu Ma, Pengjie Ren, Dawei Yin, and Zhaochun Ren.
  2023{\natexlab{a}}.
\newblock Is {ChatGPT} good at search? investigating large language models as
  re-ranking agent.
\newblock \emph{arXiv preprint arXiv:2304.09542}.

\bibitem[{Sun et~al.(2023{\natexlab{b}})Sun, Yan, Ma, Ren, Yin, and
  Ren}]{sun2023chatgpt}
Weiwei Sun, Lingyong Yan, Xinyu Ma, Pengjie Ren, Dawei Yin, and Zhaochun Ren.
  2023{\natexlab{b}}.
\newblock Is {ChatGPT} good at search? investigating large language models as
  re-ranking agent.
\newblock \emph{arXiv preprint arXiv:2304.09542}.

\bibitem[{Tang et~al.(2023)Tang, Zhang, Ma, Lin, and Ture}]{tang2023found}
Raphael Tang, Xinyu Zhang, Xueguang Ma, Jimmy Lin, and Ferhan Ture. 2023.
\newblock Found in the middle: Permutation self-consistency improves listwise
  ranking in large language models.
\newblock \emph{arXiv preprint arXiv:2310.07712}.

\bibitem[{Tay et~al.(2022{\natexlab{a}})Tay, Dehghani, Tran, Garcia, Bahri,
  Schuster, Zheng, Houlsby, and Metzler}]{ul2}
Yi~Tay, Mostafa Dehghani, Vinh~Q Tran, Xavier Garcia, Dara Bahri, Tal Schuster,
  Huaixiu~Steven Zheng, Neil Houlsby, and Donald Metzler. 2022{\natexlab{a}}.
\newblock Unifying language learning paradigms.
\newblock \emph{arXiv preprint arXiv:2205.05131}.

\bibitem[{Tay et~al.(2022{\natexlab{b}})Tay, Tran, Dehghani, Ni, Bahri, Mehta,
  Qin, Hui, Zhao, Gupta et~al.}]{tay2022transformer}
Yi~Tay, Vinh~Q Tran, Mostafa Dehghani, Jianmo Ni, Dara Bahri, Harsh Mehta, Zhen
  Qin, Kai Hui, Zhe Zhao, Jai Gupta, et~al. 2022{\natexlab{b}}.
\newblock Transformer memory as a differentiable search index.
\newblock In \emph{Advances in Neural Information Processing Systems}.

\bibitem[{Thakur et~al.(2021)Thakur, Reimers, R{\"u}ckl{\'e}, Srivastava, and
  Gurevych}]{thakur2021beir}
Nandan Thakur, Nils Reimers, Andreas R{\"u}ckl{\'e}, Abhishek Srivastava, and
  Iryna Gurevych. 2021.
\newblock {BEIR}: A heterogeneous benchmark for zero-shot evaluation of
  information retrieval models.
\newblock In \emph{Thirty-fifth Conference on Neural Information Processing
  Systems Datasets and Benchmarks Track (Round 2)}.

\bibitem[{Thomas et~al.(2023)Thomas, Spielman, Craswell, and
  Mitra}]{thomas2023large}
Paul Thomas, Seth Spielman, Nick Craswell, and Bhaskar Mitra. 2023.
\newblock Large language models can accurately predict searcher preferences.
\newblock \emph{arXiv preprint arXiv:2309.10621}.

\bibitem[{Virtanen et~al.(2020)Virtanen, Gommers, Oliphant, Haberland, Reddy,
  Cournapeau, Burovski, Peterson, Weckesser, Bright et~al.}]{virtanen2020scipy}
Pauli Virtanen, Ralf Gommers, Travis~E Oliphant, Matt Haberland, Tyler Reddy,
  David Cournapeau, Evgeni Burovski, Pearu Peterson, Warren Weckesser, Jonathan
  Bright, et~al. 2020.
\newblock Scipy 1.0: fundamental algorithms for scientific computing in python.
\newblock \emph{Nature methods}, 17(3):261--272.

\bibitem[{Wu et~al.(2023)Wu, Zheng, Qiu, Wang, Gu, Shen, Qin, Zhu, Zhu, Liu
  et~al.}]{wu2023survey}
Likang Wu, Zhi Zheng, Zhaopeng Qiu, Hao Wang, Hongchao Gu, Tingjia Shen, Chuan
  Qin, Chen Zhu, Hengshu Zhu, Qi~Liu, et~al. 2023.
\newblock A survey on large language models for recommendation.
\newblock \emph{arXiv preprint arXiv:2305.19860}.

\bibitem[{Yan et~al.(2022)Yan, Qin, Wang, Bendersky, and Najork}]{yan2022scale}
Le~Yan, Zhen Qin, Xuanhui Wang, Michael Bendersky, and Marc Najork. 2022.
\newblock Scale calibration of deep ranking models.
\newblock In \emph{Proceedings of the 28th ACM SIGKDD Conference on Knowledge
  Discovery and Data Mining}, pages 4300--4309.

\bibitem[{Zadrozny and Elkan(2001)}]{Zadrozny:Elkan:KDD01}
Bianca Zadrozny and Charles Elkan. 2001.
\newblock Learning and making decisions when costs and probabilities are both
  unknown.
\newblock In \emph{Proceedings of the 7th ACM SIGKDD International Conference
  on Knowledge Discovery and Data Mining}, page 204–213.

\bibitem[{Zadrozny and Elkan(2002)}]{Zadrozny:Elkan:KDD02}
Bianca Zadrozny and Charles Elkan. 2002.
\newblock Transforming classifier scores into accurate multiclass probability
  estimates.
\newblock In \emph{Proceedings of the 8th ACM SIGKDD International Conference
  on Knowledge Discovery and Data Mining}, page 694–699.

\bibitem[{Zhu et~al.(2023)Zhu, Yuan, Wang, Liu, Liu, Deng, Dou, and
  Wen}]{zhu2023large}
Yutao Zhu, Huaying Yuan, Shuting Wang, Jiongnan Liu, Wenhan Liu, Chenlong Deng,
  Zhicheng Dou, and Ji-Rong Wen. 2023.
\newblock Large language models for information retrieval: A survey.
\newblock \emph{arXiv preprint arXiv:2308.07107}.

\bibitem[{Zhuang et~al.(2023{\natexlab{a}})Zhuang, Qin, Hui, Wu, Yan, Wang, and
  Berdersky}]{zhuang2023beyond}
Honglei Zhuang, Zhen Qin, Kai Hui, Junru Wu, Le~Yan, Xuanhui Wang, and Michael
  Berdersky. 2023{\natexlab{a}}.
\newblock Beyond yes and no: Improving zero-shot llm rankers via scoring
  fine-grained relevance labels.
\newblock \emph{arXiv preprint arXiv:2310.14122}.

\bibitem[{Zhuang et~al.(2023{\natexlab{b}})Zhuang, Qin, Jagerman, Hui, Ma, Lu,
  Ni, Wang, and Bendersky}]{zhuang2023rankt5}
Honglei Zhuang, Zhen Qin, Rolf Jagerman, Kai Hui, Ji~Ma, Jing Lu, Jianmo Ni,
  Xuanhui Wang, and Michael Bendersky. 2023{\natexlab{b}}.
\newblock {RankT5}: Fine-tuning {T5} for text ranking with ranking losses.
\newblock In \emph{Proceedings of the 46th International ACM SIGIR Conference
  on Research and Development in Information Retrieval}, pages 2308--2313.

\end{thebibliography}
\newpage
\onecolumn

\appendix
\onecolumn
\section{Reproducibility}
\label{sec:prompt}
\subsection{Prompts for Relevance Prediction}
\label{sec:prompt_relevance}
We used the \textbf{same prompt template for all 5 datasets} evaluated in the paper. Below is the prompt template for estimating relevance in the pseudo-rater mode:
\begin{tcolorbox}
Passage: \code{\{passage\}} \\

Query: \code{\{query\}} \\

Does the passage answer the query? Output Yes or No:
\end{tcolorbox}

\subsection{Prompts for Pairwise Preference}
\label{sec:prompt_preference}
Below is the prompt template for pairwise preference in the pairwise ranking mode:
\begin{tcolorbox}
Given a query \code{\{query\}}, which of the following two passages is more relevant to the query? \\

Passage A: \code{\{passage$_1$\}} \\

Passage B: \code{\{passage$_2$\}} \\

Output Passage A or Passage B:
\end{tcolorbox}

\subsection{Code and Data Release}
Our experimental results are easily reproducible, using open-sourced LLMs and standard aggregation methods (win counting, sorting,
and sliding window) used in the work. 
We intend to release pairwise preference results on all five datasets from the two open-source LLMs to aid future research. 
Specifically, we will release the data in JSON format, which will include query-document pair information
(ids, text, label, retrieval rank and scores), along with the prompts used, the generated texts,
and relevance estimation scores.

\section{Computational Costs}
\label{sec:cost}
Our constrained regression methods are based on a traditional algorithm, the extra computation cost is negligible compared with the LLM calls. Specifically, depending on the model and the token lengths of the documents, the GPU time for LLM calls to obtain one relevance estimation or one pairwise preference could vary, but it is typically on the order of 10 ms to 1 s per LLM call. For PRP, a list of 100 documents would require at least 100 s of GPU time to obtain all pairwise preferences. The constrained regression, independent of the model or the document length, can be solved (with \code{scipy.optimize.minimize}) in about 100 ms on common CPUs for a query of 100 documents.

\section{More Results on Efficient Constrained Regression}
\label{sec:more}
\begin{table}[htbp]
\caption{Effects of initial ranker (init) and base rater (base) on different constrained regression methods on TREC-DL 2020.}
\centering
\label{tbl:result_init_base}
\begin{tabular}{ccccc}
\hline {Method} & init & base & NDCG@10 & ECE \\
\hline
\hline 
\multirow{ 2}{*}{Allpair} & - & BM25 & \textbf{0.7061} & 0.2941 \\
                          & - & PRater & 0.7054 & \textbf{0.0865} \\
\hline
\multirow{ 4}{*}{SlideWin} & {BM25} & BM25  & \textbf{0.7046} & 0.2707 \\
                           & BM25   & PRater & \textbf{0.7046} & \textbf{0.0911} \\
                           & PRater  & BM25  & 0.6939 & 0.2985 \\
                           & PRater  & PRater & 0.6939 & 0.0945 \\
\hline 
\multirow{ 4}{*}{TopAll} & {BM25} & BM25  &  0.6524 & 0.5712 \\
                         & BM25   & PRater &  0.6938 & \textbf{0.0918} \\
                         & PRater  & BM25  &  0.5949 & 0.3149 \\
                         & PRater  & PRater &  \textbf{0.7025} & 0.0966 \\
\hline
\end{tabular}
\end{table}

\subsection{LLM vs non-LLM raters}
A good relevance rater is important for the constrained regression methods to work. LLM pseudo-rater (PRater) scores are cheaper than the PRP scores, and are directly leveraged in our methods. On the other hand, BM25 scores are fast ad hoc results for result retrieval and are thus available at ranking stage. 
Here, we study the effects of replacing the LLM rater (PRater) with non-LLM rater (BM25) as the base rater for $\hat{y}$ in all constrained regression methods and as the initial ranker to select pairwise constraints in efficient sliding window (SlideWin) and top vs all pairs (TopAll) methods.

The results are summarized in \tblref{tbl:result_init_base}. We have the following observations: 
First, the choice of the base rater (Base) mainly affects the relevance prediction performance: ECE of results with PRater is significantly better than of those with BM25, as the relevance prediction performance of the constrained regression methods is mainly limited by the base scores $\hat{y}$.
In contrast, the choice of Base is nearly insignificant to the ranking performance in AllPair and SlideWin methods, but affects ranking more in the TopAll method: TopAll with PRater Base always show better NDCG than TopAll with BM25 Base.
Furthermore, the choice of the initial ranker (Init) is almost neutral on regression in terms of ECE, but has a complex effect on ranking in NDCG in SlideWin and TopAll methods. 
We note that using PRater as initial ranker in SlideWin leads to slightly worse NDCG than using BM25. This is attributable to the better alignment of LLM relevance rater and PRP ranker, so that the pairwise constraints become less informative than starting from initial ranking of BM25.
On the other hand, using PRater as initial ranker in TopAll leads to better NDCG when PRater is the base rater and worse NDCG when BM25 becomes the Base. This is attributable to the alignment of initial ranker and base rater to select useful pairwise constraints.
Based on these results, we recommend to use LLM PRater as the base rater for all constrained regression methods and use BM25 as the initial ranker for SlideWin while PRater as the initial ranker for TopAll method.

\begin{table}[htbp]
\caption{Effects of top $k$ parameters in sliding window (SlideWin) and top vs all pair (TopAll) constrained regression methods on TREC-DL 2020.}
\label{tbl:result_topk}
\centering
\begin{tabular}{ccccccHcc}
\hline
         &         & \multicolumn{4}{c}{NDCG} &  &  \\
{Method} & top $k$ & \multicolumn{1}{|c}{@1} & @5 & @10 & \multicolumn{1}{c|}{@20} & MSE & ECE \\
\hline
\hline 
\multirow{ 4}{*}{SlideWin} & 2  & \textbf{0.8580} & 0.7367 & 0.6978 & 0.6547 & 0.5454 & 0.0966 \\
                           & 5  & \textbf{0.8580} & \textbf{0.7535} & 0.7013 & \textbf{0.6698} & 0.5229 & 0.0936 \\
                           & 10 & \textbf{0.8580} & \textbf{0.7535} & \textbf{0.7046} & 0.6674 & 0.5042 & 0.0911 \\
                           & 20 & \textbf{0.8580} & \textbf{0.7535} & \textbf{0.7046} & 0.6676 & \textbf{0.4808} & \textbf{0.0890} \\
\hline 
\multirow{ 4}{*}{TopAll}  & 2  & 0.7778 & 0.7014 & 0.6762 & 0.6366 & 0.5400 & 0.0981 \\
                          & 5  & \textbf{0.8642} & 0.7319 & 0.6965 & 0.6559 & 0.5190 & \textbf{0.0954} \\
                          & 10 & 0.8549 & \textbf{0.7367} & \textbf{0.7025} & \textbf{0.6593} & 0.5405 & 0.0966 \\
                          & 20 & 0.7685 & 0.7052 & 0.6848 & 0.6520 & 0.5557 & 0.0987 \\
\hline
\end{tabular}
\end{table}

\subsection{Choice of parameter k} 
We investigate the effect of hyper-parameter $k$ in both SlideWin and TopAll methods. Note that though we have chosen the same character $k$ to represent the parameters, the actual meanings of the parameters are different in the corresponding methods: top $k$ is the number of top results to be sorted in the SlideWin, and $k$ is the number of the top results in the initial ranker to fetch pairwise constraints.

In \tblref{tbl:result_topk}, primarily, we find the choice of top $k$ affects the ranking performance (NDCGs) only. In specific, ignoring numerical fluctuations, increasing parameter $k$ of SlideWin monotonically improves NDCG@$m$ till $k\sim m$. On the other hand, increasing parameter $k$ of TopAll leads to non-monotonic NDCG@$m$ that is optimized approximately around $k\sim m$. 
The intuition of the difference between SlideWin and TopAll is that (1) the parameter $k$ of SlideWin is the top number after pairwise ordering, so that top $k$ result orders will always be consistent with PRP results so as NDCG@$m$, as long as $k>m$; 
(2) while the parameter $k$ of TopAll is the number of top results in initial ranker, which is different from the PRP results, so that when $k < m$, increasing $k$ is likely improving NDCG@$m$ as more top results are included, however, when $k > m$, more intra-top pair constraints become more dominant than top vs rest pairs, which may break the order between top $k$ vs rest results and lead to worse NDCG.

\end{document}